# High-order Harmonic Generation in Thermotropic Liquid Crystals

Andrea Annuziata[1,2,3], Luise Becker[3], Marta L. Murillo-Sánchez[3], Patrick Friebel[3], Salvatore Stagira[1,2], Davide Faccialà[2], Caterina Vozzi[2], and Laura Cattaneo[3,*]

[1]*Department of Physics, Politecnico di Milano, Milano, MI, Italy*
[2]*Istituto di Fotonica e Nanotecnologie-CNR (CNR-IFN), 20133 Milano, Italy*
[3]*Max-Planck-Institut für Kernphysik, Saupfercheckweg 1, D-69117 Heidelberg, Germany*
[*]*cattaneo@mpi-hd.mpg.de*

**Thermotropic liquid crystals are versatile optical materials that exhibit a state of matter intermediate between liquids and solids. Their properties can change significantly with temperature, pressure, or other external factors, leading to different phases. The transport properties within these materials in different phases are still largely unexplored and their understanding would enable exciting prospects for innovative technological advancements. High-order harmonic spectroscopy proved to be a powerful spectroscopic tool for investigating the electronic and nuclear dynamics in matter. Here we report the first experimental observation of high-order harmonic generation in thermotropic liquid crystals in two different phase states, nematic and isotropic. We found the harmonic emission in the nematic phase to be strongly dependent on the relative orientation of the driving field polarization with respect to the liquid crystal alignment. Specifically, the harmonic yield has a maximum when the molecules are aligned perpendicularly to the polarization of the incoming radiation. Our results establish the first step for applying high-order harmonic spectroscopy as a tool for resolving ultrafast electron dynamics in liquid crystals with unprecedented temporal and spatial resolution.**

## 1. INTRODUCTION

With the increasing need for energy efficient technologies for batteries or transistors, liquid crystals (LCs), notoriously used for displays or tunable optical elements (photonic crystals,[1] q-plates,[2,3] etc..), proved to be a promising platform for electron and ion transport combining self-assembly, stimuli response, self-healing, and easy processability.[4,5] Over the past two decades this led to a significant effort devoted to designing new LC molecules with high mobility that may become competitive candidates for organic semiconductor applications such as organic light emitting diodes, organic field effect transistors, or organic solar cells.[5–8] Although significant progress has been made in chemical synthesis and molecular design to improve transport dynamics in organic materials, some challenges have yet to be overcome for its practical implementation.[9] Among them is the lack of a detailed understanding, both experimental and theoretical, of the electronic and coupled intra-to-intermolecular dynamics upon charge transfer phenomena.[9,10]

The electronic excitation of LCs has been subject of multiple research work, in particular concerning the absorption and fluorescence activity of alkylbiphenyl-based LC molecules (nCB) when diluted in suitable solvents.[11–13] Of particular interest is the fact that these molecules within the LC mesophase pair up in excimers, when UV-excited (involving a conformational change),[14,15] and dimers in anti-parallel configuration.[16,17] The presence of such dimeric building blocks and the connected electron-nuclear dynamic, their geometrical arrangement, formation over time and connection to the LC nature is still a topic of investigation.[14–17]

In this framework, high-order harmonic generation (HHG) spectroscopy was revealed to be a valuable approach for studying electron and nuclear dynamics in multiple molecular targets.[18–22] In the well established description of HHG in gases, high-order harmonics are generated when an intense laser pulse is focused on the gas medium. Due to the strong nonlinear interaction, an electron in the medium can be freed by tunnel ionization and is accelerated by the external electric driving field. The electron may then recombine with the parent ion. The burst of light released upon recollision is composed of photons having energies going up to several tens of electronvolts with attosecond duration.[23,24] Each emitted photon energy is linked directly to a well-defined temporal interval from the ionization to the recombination of the electron, providing an extremely detailed snapshot of the structure of the system under investigation.[18,20,21,25,26] The extension of HHG spectroscopy to solids opened the way for all-optical characterization techniques such as band structure reconstruction,[27] Berry phase measurement,[28] and strong-field induced band structure dynamics.[29] Furthermore, recently Luu et al. demonstrated the possibility of generating high-order harmonics from water and most common alcohols. The generated spectra are very sensitive to the density of states and band gaps, establishing the potential of HHG spectroscopy to investigate the electronic structure in liquids.[30,31]

In the context of HHG, LCs are a special class of materials because, despite the liquid-like macroscopic appearance, they have a well-defined and controllable symmetry, orientation, and periodicity, and a low ionization potential well below 10 eV.[32] Harmonic generation in LCs in the perturbative regime has already been observed,[33,34] demonstrating the capability of non-linear spectroscopy for studying the LCs alignment and extracting nonlinear polarizability.[35,36] To the best of our knowledge, the generation of harmonics in LCs in the non-perturbative regime has not been investigated yet.

In this manuscript, we report the first experimental observation of HHG from thermotropic LCs, where phase transitions are triggered by temperature changes. In particular, we investigated nematic and isotropic mesophases. In the nematic mesophase (fig. 1a), the LC molecules lack translational symmetry, but they display macro-domains with well-defined orientational symmetry, resulting in a peaked molecular distribution oriented along a specific direction called director (fig. 1a, orange arrow). The application of an external electric field allows the orientations of all the directors of the macro-domains along the electric field lines. On the other hand, in the isotropic mesophase (fig. 1b), the molecules are randomly oriented.

We generated high-order harmonic radiation in three different LCs driving the process with a strong mid-IR laser pulse and characterizing the harmonic emission as a function of the driving field polarization direction. Our observations revealed a strong dependence of the HHG emission on the relative orientation between the driving field polarization direction and the director of the LCs molecules in the nematic phase. We show how the orientation of the director is preserved during the strong-field interaction even at intensities of the order of ~ $10^{12}$ W/cm2, which are needed for driving the HHG process, and how this orientation directly affects the efficiency of the process itself. This work constitutes a first step in understanding the ultrafast electron dynamics in LCs, elucidating the intricate interplay between the electronic and molecular dynamics associated with each specific mesophase.

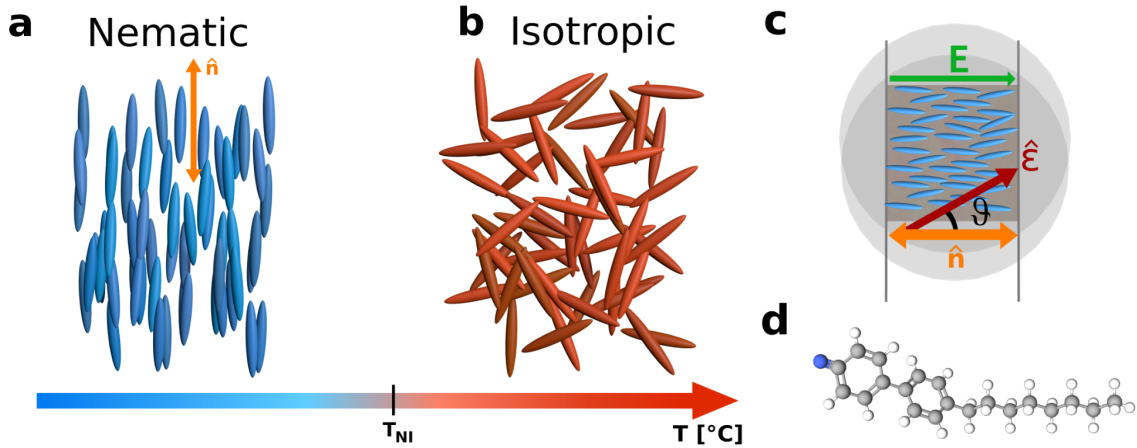

Fig. 1. **a)** LCs in the nematic phase where the molecules exhibit macrodomains characterized by a distinct orientation along an averaged direction (orange arrow), while lacking positional order. **b)** LCs in isotropic phase, where the molecules are randomly oriented. $T_{NI}$ on the arrow below indicates the nematic to isotropic transition temperature. **c)** Lab-made cell used to contain the LCs sample. The director n represents the average alignment direction (orange arrow) of the molecules and follows the external AC field. The driving field polarization direction ε is considered with respect to the director, thus with respect to the horizontal direction in the laboratory frame. **d)** Graphical depiction of the 8CB molecule. A biphenyl core with a cyanide group on one side, composed of a carbon and a nitrogen atom. The other side exhibits an 8-octyl chain terminating with a methyl group.

## 3. RESULTS AND DISCUSSION

We performed the experiments using three LCs: 8CB (4-cyano-4'-octylbiphenyl), 8OCB (4-cyano-4-n-oxyoctyl-biphenyl), and E7, which is a mixture of 5CB (4-cyano-4'-n-pentyl-biphenyl, 51%), 7CB (4-Heptyl-4-biphenylcarbonitrile, 25%), 8OCB (16%), and 5CT (4-cyano-4-n-pentyl-p-terphenyl, 8%). LCs were exposed to the laser beam (P-polarized, centered at the wavelength of 3.8 μm) either as contained in lab-made cells (sketch in Fig. 1c) or free-standing thin films. In this case, we only used 8CB among the three LCs here investigated since it is the only one providing a sufficient surface tension to form a stable film over time. Details about the experimental setup, free standing film formation, the cell preparation and temperature stabilization can be found in the supplementary material (SM).

We first performed an intensity scan in a free-standing film of 8CB at room temperature (20 °C) to assess the non-perturbative nature of the observed harmonic emission. At this temperature, 8CB is in the smectic A phase. Here, the molecules maintain the general order observed in the nematic phase but also align themselves in distinct layers. We kept this temperature because the 8CB film formed in the smectic A phase exhibits greater stability compared to the nematic phase. This ensures the preservation of the molecules' ability to pair up in dimers, without compromising the local order probed by the HHG mechanism.

Figure 2 shows the intensity of the 5$^{th}$ (a) and 7$^{th}$ (b) harmonic as a function of the driving field intensity. The green (blue) triangles represent the experimental data obtained for H5 (H7), whereas the dashed lines show the expected behavior in perturbative regime. For a driving field intensity up to 16 TW/cm$^2$ the harmonics scaling is clearly perturbative, while above this value a saturation of the emitted harmonics is observed, indicating the non-perturbative origin of the HHG process.

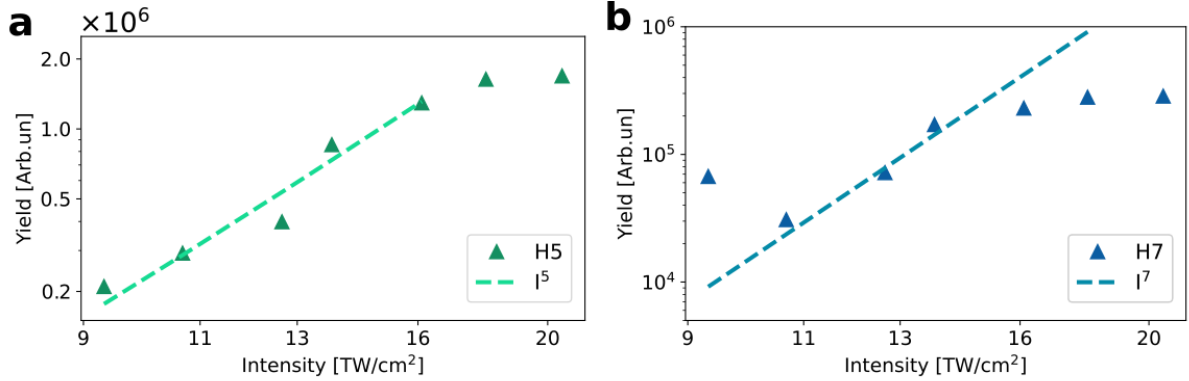

Fig. 2. Yield of the 5th harmonic a) and the 7th harmonic b) as a function of the driving field intensity for a thin film of 8CB. Both axes are in $\log_{10}$ scale. The experimental data is represented by triangles, whereas the dashed line represents the power law at the same order as the emitted harmonic.

To perform HHG experiments across all LC mesophases and different alignment configurations, the LC-filled cell was preferred compared to the free-standing film since the latter, in fact, survives only within a limited range of temperature (below 33 °C) and LC orientation (molecules arranged only perpendicular to the film surface).

Moving to the LC cells, we analyzed the harmonic emission from E7, 8CB, and 8OCB in the nematic phase, by setting the temperature below the transition temperature to the isotropic phase (60 °C, 40.5 °C and 80 °C, respectively).

For all the LCs under investigation, the harmonics were collected as a function of the relative orientation between the applied external field and the (linear) polarization direction of the driving field. This orientation was identified by the angle θ, which was varied using a half waveplate, in steps of 8°.

The laser intensity was tuned to avoid unwanted phase transitions and degradation of the sample during the measurement. At the focus, the peak intensity of the driving field was estimated on the order of 26 TW/cm$^2$ for E7, 21 TW/cm$^2$ for 8CB, and 18 TW/cm$^2$, for 8OCB.

Figure 3 shows the yield of the detected harmonics as a function of the laser polarization direction θ for (a) E7, (b) 8CB, and (c) 8OCB. For θ= 0°, 180° the driving field is aligned parallel to the AC field, while for θ = ± 90° the driving field is aligned perpendicular to the AC field. We refer to these two cases as parallel and perpendicular configurations, respectively. For all the harmonics, the yield is normalized to the maximum. For all the LCs under investigation, only odd-order harmonics were detected. For E7 and 8CB, the 3rd, 5th and 7th harmonic (H3, H5, and H7) were observed during the polarization scan. In 8OCB only H3 and H5 were collected, because of the lower driving field intensity used. In all three samples, all the harmonics showed maximum yield in the perpendicular configuration (θ = ± 90°), giving rise to a 2-fold symmetry in the polar plots. The yield then quickly drops as we depart from the perpendicular configuration, producing two distinctive lobes. In particular, we observe that the width of these lobes decreases by increasing the harmonic order for all the LCs under investigation. Moreover, distinct weaker lobes are observed for H3 in the parallel configuration (θ = 0°, 180°). The intensity of these weak lobes relative to the main ones is 8.5%, 15%, and 18% for E7, 8CB, and 8OCB, respectively. For higher-order harmonics the harmonic emission in the parallel configuration drastically drops to the background level, resulting in a non-detectable intensity.

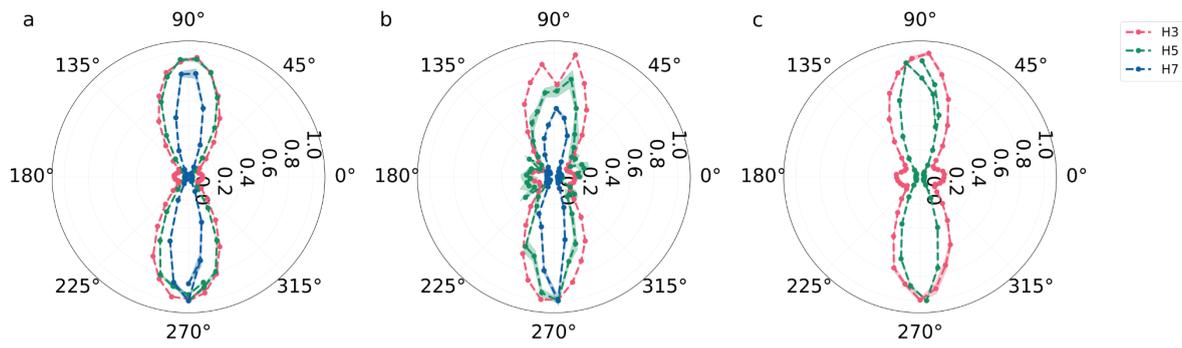

*Fig. 3 Normalized intensity of the emitted harmonics in the nematic phase for (a) E7,(b) 8CB, and (c) 8OCB as a function of the polarization direction at 36 °C, 36 °C, and 74 °C, respectively.*

For each LC under study, right after measuring the nematic phase, we switched off the applied voltage and we heated the sample above the nematic-to-isotropic transition temperature, and characterized the HHG emission in the isotropic phase, keeping the same driving field intensity as in the nematic phase. Figure 4 shows the HHG response in (a) E7, (b) 8CB, and (c) 8OCB as a function of the laser polarization direction in the isotropic phase at temperatures of 80 °C, 90 °C, and 100 °C, respectively. We expect LC molecules to be randomly aligned and therefore the polarization direction of the driving field should not affect the efficiency of the HHG process. The results show indeed a weaker dependence of the HHG intensity with the polarization angle, particularly for H3. A residual modulation can be observed for higher orders, even if not as pronounced as in the nematic phase, with a preferential emission in the perpendicular configuration. We attribute this residual modulation to the persistence of partial alignment induced by the contact with the substrate. Indeed, even in the isotropic phase, surface effects dominate, favoring a specific anchoring of the neighboring molecules to the $CaF_2$ substrate which imposes a residual alignment. This in turn modulates the emitted harmonics as a function of the driving field polarization direction.

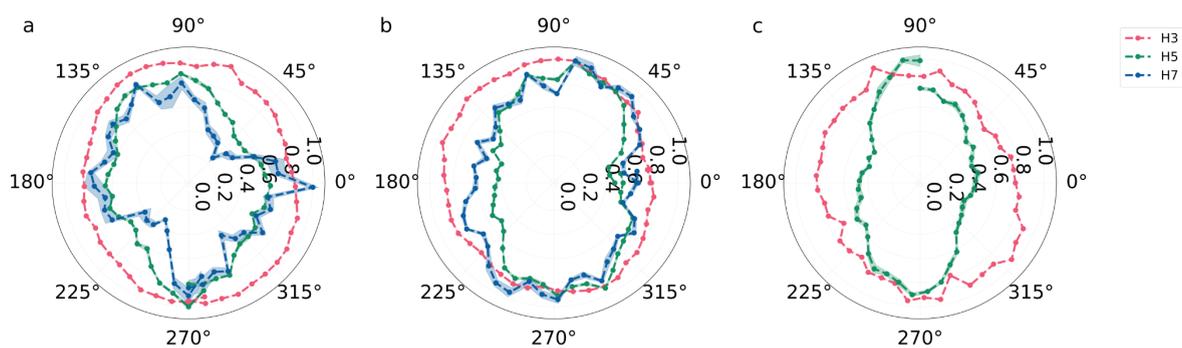

*Fig. 4. Normalized intensity of the emitted harmonics in the isotropic phase for (a) E7, (b) 8CB, and (c) 8OCB as a function of the polarization direction at 80 °C, 90 °C, and 100 °C, respectively*

To further investigate the origin of the high yield obtained in the nematic phase aligned in the perpendicular configuration, we compared the HHG spectra obtained in the two phases in this configuration. We focused our attention on E7, which, being specifically designed for display applications, is the most stable in terms of nematic-temperature range (-60 to 60 °C) and shows a more efficient alignment with the external AC field (no surface treatments were used in the prepared LC cells). Figure 5 presents a comparison of the spectra in both cases, along with the overall enhancement observed when transitioning from the isotropic to the aligned nematic phase. The harmonics generated in the aligned nematic phase are significantly more intense than the ones obtained in the isotropic phase, with the harmonic yield monotonically increasing as a function of the harmonic order. In the nematic phase, a red shift is also observed compared to the isotropic phase for all the harmonics. Specifically, the shifts are 10 nm for the H3, 13 nm for H5, and 10 nm for H7. The origin of this redshift may be attributed to a non-negligible nuclear motion during the HHG process.[39,40] This phenomenon is usually related to higher ionization rates and smaller ionization potential under a photo-dissociation process[39,41–43] or to intermediate transitions during the HHG.[44] To properly understand this observation, numerical calculations are needed.

Next, we discuss the origin of the high HHG yield observed in the nematic phase compared to the isotropic phase. This could be attributed to several effects. If we make the assumption that the HHG emission comes only from the single-molecule response, and we disregard macroscopic phase-matching effects, we expect a higher yield from the aligned nematic phase in the perpendicular configuration as compared to the isotropic phase. Indeed in the aligned nematic phase, all molecules are aligned along the axis that maximizes HHG, and they all contribute to the HHG process with the same yield. On the other hand, in the isotropic phase, molecules are randomly oriented with respect to the driving field polarization direction, resulting in a reduced or negligible contribution from molecules that are not aligned perpendicularly to the driving field. To understand the relative impact of this alignment effect on the observed enhancement, we averaged the HHG emission over all the possible orientations of the driving field laser polarization in both nematic and isotropic phases. Calling $I_N(\theta)$ the detected Nth harmonic intensity for a driving field polarization direction $\theta$, the average yield over all the possible three-dimensional laser field orientations is given by:

$$\int_0^\pi d\theta\, I_N(\theta) \sin(\theta)/2.$$

If the alignment effect is responsible for the enhancement of the harmonic emission in perpendicular configuration, the ratio between the average nematic and isotropic HHG yield should be equal to one. The calculated average ratio for all the harmonics observed is shown with green circles (right axis) in Fig. 5. For H3, the experimental results suggest that the enhancement can be entirely attributed to the alignment effect. However, for H5 and H7, the alignment effect alone cannot account for the observed enhancement. The enhancement of the average HHG yield must then originate from either the macroscopic collective coherent response of the aligned LCs or from a change in the microscopic HHG response in the two different phases of the LCs.

As previously mentioned, it has been identified that nCB (and nCOB[12]) molecules form dimeric clusters $(nCB)_2$ which imposes to the involved molecular units a reduced intermolecular distance (below 4 Å) and a defined antiparallel configuration in diluted solution of different solvents.[14–17] In particular the dimeric contribution becomes dominant at high concentrations up to the neat phase.[11–13,16] When looking at the fluorescence spectra of pure 8CB at different temperatures, the dimer/excimer contribution is dominant in the nematic and smectic phase, it becomes equally shared with the monomeric contribution in the isotropic phase and almost negligible in the crystalline phase where almost pure monomeric emission is observed.[11] In our case, we could think that the observed enhanced emission in the nematic phase is the result of two factors: i) the presence of dimeric clusters favoring a larger delocalization of the excited electron wave-packet thanks to the coupling of the two bi-phenyl groups almost irrespective of the alkyl/alkoxy chain; ii) the alignment over the entire bulk of such dimeric units which amplify the efficiency of the triggered process, in particular this happens when the incoming radiation is polarized perpendicular to the LC director but most probably along the dimeric highest occupied molecular orbital (HOMO). On the other hand, in the isotropic phase the reduced and isotropic response can be attributed to the reduced concentration of dimeric clusters and their lack of alignment.

To find evidence of our interpretation, it is essential to conduct a direct quantitative comparison between our experimental results and theoretical calculations. This would allow us to clarify the mechanism behind the non-perturbative HHG emission and to identify possible additional transport mechanisms involving hopping/delocalization among neighboring dimers. Finally, phase-matching could also contribute as an additional possible source of the observed average yield enhancement.

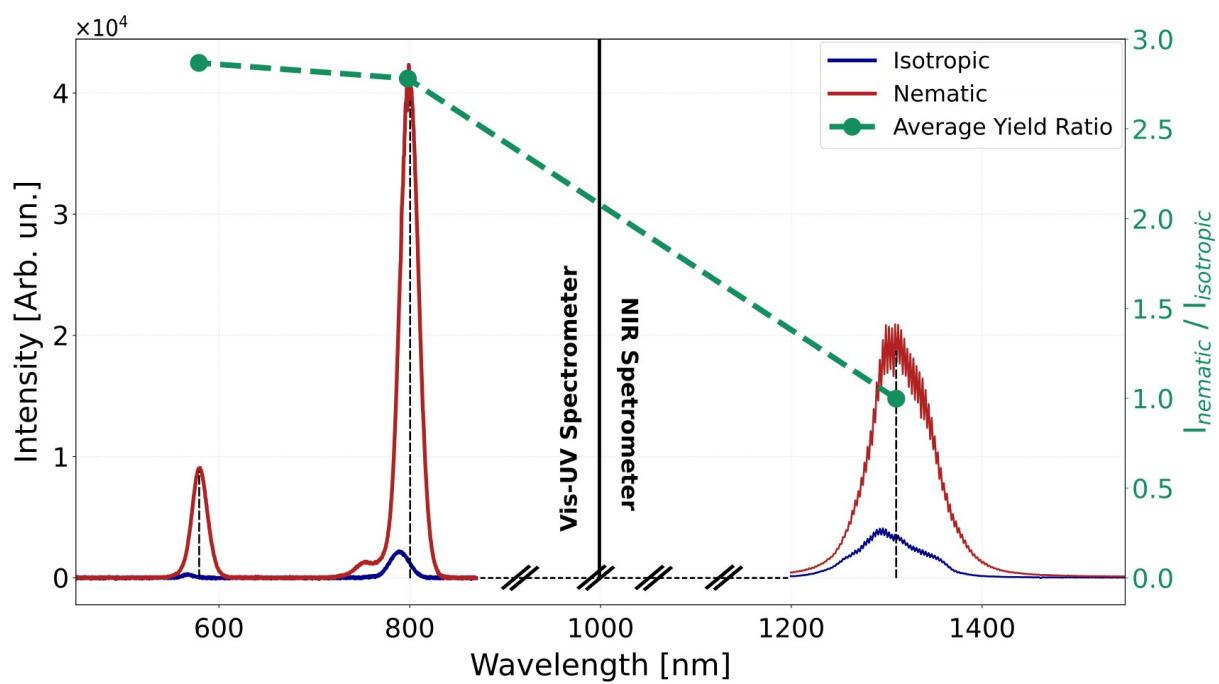

FIG. 5. Harmonic response in the isotropic (blue) and nematic (red) phases for the same polarization direction of the driving field in E7. The harmonics in the visible and near-infrared ranges were acquired with two different spectrometers. Due to the absence of overlapping regions, it is not possible to compare the relative intensities between the NIR and VIS-UV ranges. Green circles (right axis) represent the total yield ratio averaged over all possible polarization directions with respect to the molecular alignment between the measured intensity for the nematic and isotropic phase. A red-shift for all the harmonic order is also observed, represented by black dashed lines.

## 4. CONCLUSION

In this study, we report the first experimental observation of high-order harmonic generation in the non-perturbative regime from three different thermotropic LCs. The harmonic emission as a function of the driving field polarization direction has been characterized within both the aligned nematic and isotropic mesophases.

Our experimental observation reveals a strong dependence of the harmonic emission on the LC phase. Indeed, a strong modulation of the harmonic emission was observed by changing the polarization direction of the driving field in the nematic phase, with a strong enhancement when the driving field was orthogonally polarized to the director and a clear suppression when the driving field was parallel. In contrast, the harmonic response in the isotropic phase exhibited a more uniform dependence on the driving field polarization direction.

Moreover, the harmonic yield showed a significant enhancement in the nematic phase compared to the isotropic phase. Our analysis showed that the observed enhancement cannot be solely attributed to the molecular alignment for the harmonics above the 3rd order. We argue this result considering the presence of aligned anti-parallel dimeric clusters. The dimeric configuration favors the delocalization of the electronic wavefunction around the coupled biphenyl groups. In particular the observed enhanced HHG emission indicates a perpendicular spatial distribution of the HOMO wave-function compared to the molecular/dimer orientation. The overall alignment of all dimers within the bulk increases the efficiency of the HHG process fixing the relative geometry between the polarization of the incoming radiation and the emitting units. In conclusion, HHG is confirmed to be a powerful investigation tool sensitive to the local order in a soft matter system. Moreover, it is particularly adapt at identifying the critical factors that lead to the formation of delocalized states. This is a critical aspect for understanding long-range transport processes in soft matter and harnessing them for future technological applications. This work represents a first step in understanding the ultrafast dynamics in LCs, unfolding the complexity of the electron and molecular dynamics ruling each mesophase at unprecedented spatial and time scales. Moreover, since their temperature-controlled mesophase nature, thus the possibility to go from an oriented and periodic structure to a fully random system, LCs will serve as ideal benchmark materials to bridge molecular and solid-state high-order harmonic generation exploiting phase transitions.

**FUNDING**

This publication is based upon work from COST Action Attochem CA18222, supported by COST (European Cooperation in Science and Technology).

C.V., D.F. and A.A. acknowledges funding through the MIUR PRIN CONQUEST (Grant No. 2020JZ5N9M).

**ACKNOWLEDGMENTS**

L.C. would like to acknowledge the Max Planck Group Leader program for funding her independent research.

## DISCLOSURES

The authors declare no conflicts of interest.

## DATA AVAILABILITY

The data that support the findings of this study are available from the corresponding author upon reasonable request.